\documentclass{IEEEtran}
\usepackage{cite}
\usepackage{amsmath,amssymb,amsfonts}
\usepackage{graphicx}
\usepackage{subcaption} 
\usepackage{xcolor} 
\usepackage{textcomp,nicefrac}
\usepackage{algorithm}
\usepackage{algpseudocode}
\usepackage{array}
\usepackage{booktabs}
\def\BibTeX{{\rm B\kern-.05em{\sc i\kern-.025em b}\kern-.08em
T\kern-.1667em\lower.7ex\hbox{E}\kern-.125emX}}
\begin{document}
\title{Jet Image Generation in High Energy Physics Using Diffusion Models}
   
\author{Victor D. Martinez, Vidya Manian, ~\IEEEmembership{Senior Member,~IEEE,} and Sudhir Malik
\thanks{This paragraph of the first footnote will contain the
date on which you submitted your paper for review. It will also contain
support information, including sponsor and financial support acknowledgment.
\\This work was supported by the U.S. National Science Foundation under Grant 2334265.
 
Victor Diaz-Martinez is with the Department of Electrical and Computer Engineering, University of Puerto Rico, Mayaguez, PR 00681, USA. (e-mail: victor.diaz16@upr.edu).}
\thanks{Vidya Manian, is with the Department of Electrical and Computer Engineering and Bioengineering, University of Puerto Rico, Mayaguez, PR 00681 USA. (e-mail: vidya.manian@upr.edu).}
\thanks{Sudhir Malik, is with the Department of Physics, University of Puerto Rico, Mayaguez, PR 00681 USA. (e-mail: sudhir.malik@upr.edu).
}
} 

\maketitle

\begin{abstract}
This article presents, for the first time, the application of diffusion models for generating jet images corresponding to proton-proton collision events at the Large Hadron Collider (LHC). The kinematic variables of quark, gluon, W-boson, Z-boson, and top quark jets from the JetNet simulation dataset are mapped to two-dimensional image representations. Diffusion models are trained on these images to learn the spatial distribution of jet constituents. We compare the performance of score-based diffusion models and consistency models in accurately generating class-conditional jet images. Unlike approaches based on latent distributions, our method operates directly in image space. The fidelity of the generated images is evaluated using several metrics, including the Fréchet Inception Distance (FID), which demonstrates that consistency models achieve higher fidelity and generation stability compared to score-based diffusion models. These advancements offer significant improvements in computational efficiency and generation accuracy, providing valuable tools for High Energy Physics (HEP) research.

\end{abstract}


\section{Introduction} \label{sec
}

\IEEEPARstart{D}{iffusion} models have been used for a wide range of image generation tasks, including grayscale images, RGB color images, hyperspectral images, and physics-based images. Grayscale and color image generation using diffusion models have demonstrated significant advancements in capturing details and color distributions. In grayscale image generation, these models effectively reproduce variations in intensity and texture, as shown in recent studies \cite{ho2020denoising, nichol2021improved}. For color images, diffusion models handle the complexity of color channel interactions, resulting in realistic and vibrant outputs that rival those produced by traditional generative models \cite{dhariwal2021diffusion}. Techniques such as improved neural network architectures and attention mechanisms have further enhanced the quality and diversity of generated images \cite{nichol2021improved}. 

These models operate by iteratively refining random noise through a diffusion process, which reverses a series of small, incremental degradations. This approach allows them to capture complex data distributions and generate realistic images, often outperforming previous generative models like Generative Adversarial Networks (GANs) \cite{ho2020denoising}. The core idea behind diffusion models is to start with a sample of pure noise and gradually transform it into a coherent image through a series of denoising steps. This process is defined by a Markov chain, where each step involves predicting and removing noise using neural networks trained on large datasets. The training phase involves learning to reverse the diffusion process, essentially teaching the model to reconstruct data by learning the probability distribution of the training data. This step-by-step refinement allows diffusion models to generate images with finer details and fewer artifacts compared to traditional models \cite{sohl2015deep}.

The LHC is the world's largest and most powerful particle accelerator, located at the CERN laboratory near Geneva, Switzerland. At the LHC, proton beams collide at extremely high energies, producing stable and unstable particles that are detected and analyzed to investigate the fundamental laws of physics and to search for phenomena beyond the Standard Model.

Two primary detectors at the LHC, ATLAS \cite{atlas2023} and CMS\cite{cms2023}, are general-purpose instruments designed to capture a wide range of physical phenomena arising from these collisions. As protons collide with protons, particles split at locations called primary vertices producing stable particles such as quarks and gluons which have a certain amount of energy as they travel further into the collider layers \cite{cms2008cms}. These particles reach a stage called hadronization and break up or degrade into showers of unstable particles at locations called secondary vertices \cite{Shlomi_2021}. These decays often involve heavy-flavor particles like b-quarks and c-quarks, which have relatively long lifetimes and travel a measurable distance before decaying. The detectors record the events that produce these so called "jets," collimated sprays of secondary particles resulting from the fragmentation of initial quarks and gluons via the strong interaction. Detailed analysis of these jets enables researchers to explore complex physical processes, including the identification of new particles and potential signs of unknown interactions. These jets consist of numerous secondary particles that emerge from the decay of primary particles, creating dense and highly variable spatial distributions \cite{cms2023, atlas2023}.  Accurate representation of these jets is essential for studying fundamental processes in particle physics, including the identification of new particles and the exploration of phenomena beyond the standard model \cite{gomez2021generative}.

For jet physics analysis, the jet representation plays an important role. While sequences of particles or trees have been traditionally used, the recent mode of representing jets as point cloud of unordered set of particles has gained predominance due to its permutation invariance property \cite{Komiske_2019}. Jet tagging is the method of labeling a jet as originating from a particle of certain class. Several machine learning (ML) methods have been developed for tagging jets, such as the particle transformer \cite{qu2024particletransformerjettagging}. ParticleNet \cite{particleNet} and particle transformer \cite{wu2024jettaggingmoreinteractionparticle} are other architectures developed for jet tagging.

Diffusion models have been applied across various domains, including image super-resolution, inpainting, and conditional image generation. Recent advancements have made these models faster and more efficient, addressing one of their main drawbacks, computational time. Techniques such as improved training procedures, model optimizations, and hybrid approaches combining elements from other generative models have significantly enhanced their performance and practicality in real-world applications. As a result, diffusion models are rapidly becoming a leading choice for image generation tasks in both research and industry \cite{dhariwal2021diffusion}.\\


\noindent \textbf{Diffusion Models for Jet Generation:}
Diffusion models are extensively used at different stages of the collider for data generation. Score based Diffusion models are built on Stochastic Differential Equations (SDE). Generative models generate new observations from a noisy distribution. Score-based generative models (SGMs) are a class of generative model that learns to map noise to data by estimating the score function (the gradient of the log-probability density) of the data distribution. This approach avoids the computationally expensive task of calculating the partition function, which is required by many other generative models. The calorimeter shower refers to the cascading process in which a high-energy particle (like an electron or photon) interacts with the calorimeter material, generating a shower of secondary particles. SDE models have been employed to generate calorimeter shower images \cite{Mikuni_2022}. Liquid Argon Time Projection Chamber particle trajectories have also been generated with SDE method \cite{imani2024scorebaseddiffusionmodelsgenerating}, \cite{shuchin_paper}. While the above two applications have been for particle image generation, majority of the diffusion models have been applied for generation of point cloud jets. Equivariant Generative Adversarial Networks \cite{Epic_GAN} and transformers \cite{diffusion_GPT} have been used for point cloud jet generation. Jets are represented using the constituent particle kinematic variables of $(p_T, \eta, \phi)$, where $p_T$ is the transverse momentum, $\phi$ is the azimuthal angle measured with respect to the $x$-axis, and $\eta$ is the pseudorapidity. Each jet can have number of particles ranging from 25 to 100. The original JetNet dataset consisting of jets in point cloud representation was created based on the principle of Message Passing Generative Adversarial (MPGAN) neural network \cite{Kansal_JetNet_2023} and each jet consists of 25 to 30 particles. Using JetNet jet data as input, jets have been generated in point cloud form using score-based diffusion models \cite{PhysRevD.108.036025}, PC-JEDI model \cite{Leigh_2024jedi} and consistency PC-DROID model \cite{Leigh_2024}.  
\\

\noindent \textbf{Jet Image Generation Methods:} 
In image based generation, each point cloud jet with constituent particle kinematics simulated using Pythia \cite{Komiske_2019} are represented in a rectangular 2D grid as images.  The oldest method for jet image generation was done using a location-aware Generative Adversarial Network (LAGAN) that consists of a generator and a discriminator block \cite{de_Oliveira_2016}. This method was used to generate boosted W boson jet images. A Variational Autoencoder  (VAE) was implemented to generate the same jet images \cite{Dohi:2020eda}. A score-based diffusion model was distilled into a consistency model for generation of calorimeter shower simulation images in \cite{buhmann2024calocloudsiiultrafastgeometryindependent}. Generating jet images of particle jets involves several challenges \cite{qi2017pointnet}, such as the sparsity of the images compared to natural images. Unlike natural images from commonly studied datasets, jet images have very low information and are highly sparse (number of pixels with nonzero values $\sim$ 10 to 20\%). Jet particles often vary in density depending on the class of jet image generated demanding models that can handle variable-sized inputs while maintaining scalability \cite{batra2018gan}. Ensuring physical realism in the generated images, such as maintaining accurate energy and momentum distributions, is also crucial \cite{gomez2021generative}. Furthermore, evaluating the quality of generated images requires specialized metrics  that assess both visual fidelity and physical accuracy \cite{barron2018deep}. In this work, we apply diffusion models to generate five classes of jet images from the JetNet dataset. The dataset and its preprocessing are described in detail in Section III.C.

This article presents SGM and consistency models for generation of jet images, offering an efficient and accurate alternative to previous point cloud jet generation  methods. The main contributions of this research are summarized as follows:

\begin{itemize} \item Implementation of score-based models for the generation of jet images trained on JetNet dataset. \item Development of consistency models for the generation of jet images trained on JetNet dataset. 
\item Reconstruction of jet mass from generated images and comparison of the two methods for jet image generation with original jet mass.
\item Evaluation of the jet image generations using several metrics to assess the fidelity and accuracy of the generated jet images.
\item Statistical significance analysis of jet image generation using diffusion models. \end{itemize}

The remainder of this paper is organized as follows: Section II provides an overview of diffusion models, including score-based and consistency models. Section III details the jet image generation experiments including JetNet dataset used for training the diffusion models, the experimental setup and reconstruction of particles from generated jet images. Section IV presents the results of the jet image generation experiments, followed by a discussion in Section V. Finally, Section VI concludes the paper and outlines potential future work.



\section{Diffusion models}
Diffusion models constitute a class of probabilistic generative models that initially alter data by gradually adding noise. Subsequently, these models learn to reverse this process, enabling the generation of new samples from noisy data. Currently, research in this field focuses on three main approaches: denoising diffusion probabilistic models (DDPMs) \cite{sohl2015deep,ho2020denoising}, Score-based Generative Models (SGMs) \cite{song2019generative,song2020improved}, and Stochastic Differential Equations (Score SDEs) \cite{song2021maximum,song2020score}.

\subsection{Score-Based Generative Models}

\begin{figure}[htbp]
    \centering
    \includegraphics[width=0.8\columnwidth]{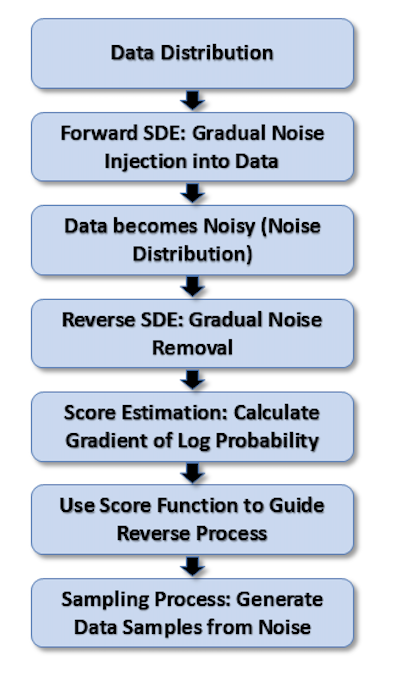}
    \caption{Score-based generative modeling: The process begins with data distribution, where noise is gradually injected through a Forward SDE. The data becomes noisy, and the reverse process uses the Reverse SDE to remove noise, estimating the score and generating samples from the noise.} 
    \label{fig:score_based_generative_modeling}
\end{figure}

At the heart of SGMs \cite{song2019generative, song2020improved} lies the concept of gradually adding noise to data, and then reversing the process using SDEs. This process relies on the Stein score (also known as the score or score function) \cite{hyvarinen2005estimation}, which represents the gradient of the logarithm of the probability density of the data. The diagram in Figure \ref{fig:score_based_generative_modeling} illustrates the key steps involved in this process, where noise is progressively injected into the data, followed by score estimation and the reverse SDE to generate new samples.

\begin{algorithm}[h!]
\caption{Score-Based Generative Modeling for Jet Images (VE SDE)}
\label{alg:score_sde}
\begin{algorithmic}[1]
\Statex \textbf{Part 1: Training the Score Network}
\State \textbf{Require:} Jet image dataset $\mathcal{D}$, score network $s_\theta(x, t)$, time horizon $[0, T]$.
\For{each training iteration}
    \State Sample a jet image $x_0 \sim \mathcal{D}$.
    \State Sample a time $t \sim \mathcal{U}(0, T)$.
    \State Sample Gaussian noise $z \sim \mathcal{N}(0, \mathbf{I})$.
    \State Compute the standard deviation $\sigma_t$ for time $t$.
    \State Perturb the data: $x_t = x_0 + \sigma_t z$.
    \State Compute the loss: $\mathcal{L}(\theta) = \left\| s_\theta(x_t, t) + \frac{z}{\sigma_t} \right\|_2^2$.
    \State Update model parameters $\theta$ using a gradient descent step on $\mathcal{L}(\theta)$.
\EndFor
\Statex
\Statex \textbf{Part 2: Sampling with Predictor-Corrector (PC)}
\State \textbf{Require:} Trained score network $s_\theta$, noise scales $\{\sigma_i\}_{i=1}^N$, Langevin steps $M$, SNR $r$.
\State Sample initial noise $x_N \sim \mathcal{N}(0, \sigma_N^2 \mathbf{I})$.
\For{$i = N, \dots, 1$}
    \Statex \Comment{\textit{Corrector: Refine sample with Langevin MCMC}}
    \For{$j = 1, \dots, M$}
        \State Sample $z \sim \mathcal{N}(0, \mathbf{I})$.
        \State Let $g \leftarrow s_\theta(x_i, \sigma_i)$.
        \State Set step size $\epsilon \leftarrow 2(r \sigma_i)^2$.
        \State $x_i \leftarrow x_i + \epsilon g + \sqrt{2\epsilon} z$.
    \EndFor
    \Statex \Comment{\textit{Predictor: Solve the reverse SDE for one step}}
    \State Sample $z \sim \mathcal{N}(0, \mathbf{I})$ if $i > 1$, else $z=0$.
    \State $x_{i-1} \leftarrow x_i + (\sigma_i^2 - \sigma_{i-1}^2) s_\theta(x_i, \sigma_i) + \sqrt{\sigma_i^2 - \sigma_{i-1}^2} z$.
\EndFor
\State \textbf{return} $x_0$.
\end{algorithmic}
\end{algorithm}

 Given a probability density function $p(\mathbf{x})$, the score function is defined as the gradient of the logarithm of the probability density, i.e., $\nabla_{\mathbf{x}} \log p(\mathbf{x})$. Unlike the Fisher score $\nabla_{\theta} \log p_{\theta}(\mathbf{x})$ commonly used in statistics, the Stein score considered here is a function of the data $\mathbf{x}$ rather than the model parameters $\theta$. It is a vector field that points in the directions where the probability density function exhibits the highest growth rate.

The core idea of SGMs \cite{song2019generative} is to perturb data with a sequence of progressively stronger Gaussian noise levels and jointly estimate the score functions for all noisy data distributions by training a deep neural network model conditioned on noise levels (referred to as a noise-conditional score network, NCSN, in \cite{song2019generative}). Samples are generated by linking the score functions at decreasing noise levels using score-based sampling approaches, such as Langevin Monte Carlo \cite{ song2019generative}, stochastic differential equations \cite{song2020score}, ordinary differential equations \cite{song2020improved}, and various combinations thereof \cite{dhariwal2021diffusion}. In SGMs, the processes of training and sampling are fully decoupled, allowing for the application of a wide range of sampling techniques once the score functions have been estimated.

Let $q(\mathbf{x}_0)$ represent the data distribution, and let $0 < \sigma_1 < \sigma_2 < \dots < \sigma_t < \dots < \sigma_T$ denote a sequence of noise levels. A typical SGM example involves perturbing a data point $\mathbf{x}_0$ to $\mathbf{x}_t$ using a Gaussian noise distribution $q(\mathbf{x}_t \mid \mathbf{x}_0) = \mathcal{N}(\mathbf{x}_t; \mathbf{x}_0, \sigma_t^2 I)$. This process results in a sequence of noisy data densities $q(\mathbf{x}_1), q(\mathbf{x}_2), \dots, q(\mathbf{x}_T)$, where
\begin{equation}
    q(\mathbf{x}_t) := \int q(\mathbf{x}_t \mid \mathbf{x}_0)\, q(\mathbf{x}_0)\, d\mathbf{x}_0.
\end{equation}
A noise-conditional score network is a deep neural network $s_\theta(\mathbf{x}, t)$ trained to estimate the score function $\nabla_{\mathbf{x}_t} \log q(\mathbf{x}_t)$. Learning score functions from data (i.e., score estimation) has been established through techniques such as score matching \cite{hyvarinen2005estimation}, denoising score matching \cite{vincent2011connection}, and sliced score matching \cite{kolouri2019sliced}, allowing us to directly train our noise-conditional score networks using perturbed data points. Figure 2 shows the Forward and Reverse SDE process. Algorithm 1 gives the pseudocode for Variance Exploding (VE) SDE training, wherein the variance of the noise added to the original image during the forward diffusion process increases over time. The algorithm gives the steps for training the score network and sampling with Langevin Markov Chain Monte Carlo (MCMC) for generating new denoised jet image data samples.

\begin{figure}[htbp]
    \centering
    \begin{subfigure}[b]{\columnwidth}  
        \centering
        \includegraphics[width=\textwidth]{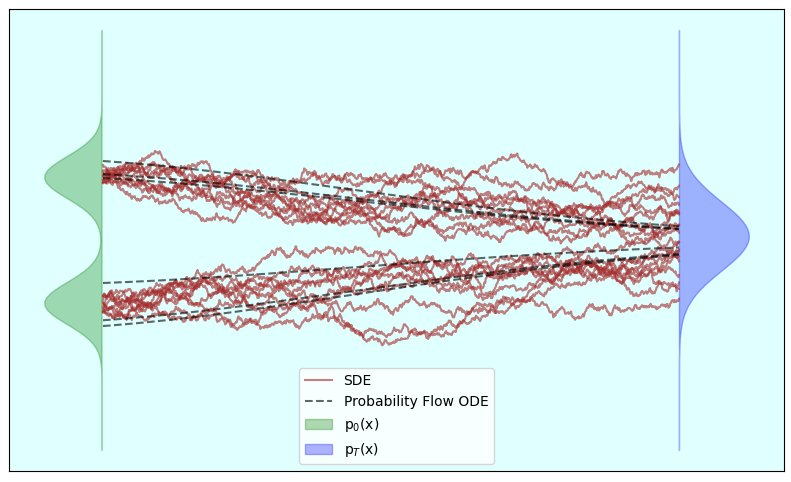}
        \caption{ Forward SDE.}
        \label{fig1a} 
    \end{subfigure}
    \par\medskip  
    \begin{subfigure}[b]{\columnwidth} 
        \centering
        \includegraphics[width=\textwidth]{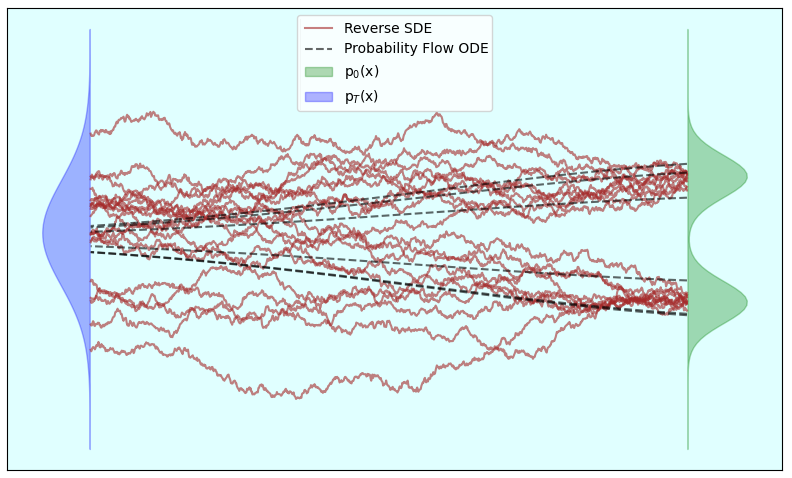}
        \caption{Reverse SDE.}
        \label{fig1b}
    \end{subfigure}
    \caption{Overview of score-based generative modeling through SDEs. In (a), data can be mapped to a noise distribution (the prior) using a forward SDE, and in (b), this process is reversed for generative modeling using a reverse SDE. The associated probability flow ODE in (b) can also be reversed, yielding a deterministic process that samples from the same distribution as the forward SDE. Both the reverse-time SDE and probability flow ODE are derived by estimating the score $\nabla_x \log p_t(x)$. 
}
    \label{fig1}
\end{figure}

\subsection{Consistency Models}

\begin{figure}[htbp]
    \centering
    \includegraphics[width=0.8\columnwidth]{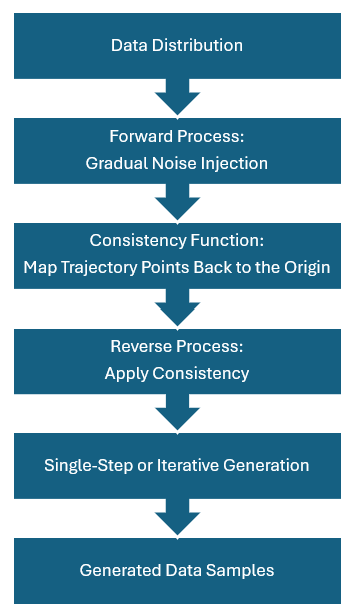}
    \caption{Steps of the consistency model.} 
    \label{fig:consistency_steps}
\end{figure}

Consistency models are based on the foundational work in diffusion models and score-based generative modeling. In particular, Song et al. \cite{song2023consistency} introduced the concept of consistency functions and demonstrated their efficacy in generating high-quality images with reduced computational costs. Subsequent studies have expanded on these ideas, exploring different training regimes and applications in various domains, including physics-based simulations \cite{gomez2021generative} and image synthesis \cite{nichol2021improved}.\\
In the context of HEP, consistency models offer promising avenues for generating jet images. They have been used for point cloud jet generation \cite{Leigh_2024} and calorimeter point cloud shower generation\cite{buhmann2024calocloudsiiultrafastgeometryindependent}. Their ability to produce high-fidelity samples in a single step makes them particularly suitable for large-scale simulations required in experiments such as those conducted at the LHC. Moreover, the single or few step generation capability can be leveraged to simulate various physical scenarios without the need to retrain the model, thus improving the flexibility and efficiency of simulation workflows \cite{song2023consistency, gomez2021generative}. The process of consistency models for jet image generation can be summarized in several key steps, as illustrated in Figure \ref{fig:consistency_steps}.


\begin{algorithm}[h!]
\caption{Consistency Training (CT) for Jet Images}
\label{alg:consistency_training}
\begin{algorithmic}[1]
\Statex \textbf{Part 1: Training the Consistency Model}
\State \textbf{Require:} Jet image dataset $\mathcal{D}$, online model $f_\theta$, target model $f_{\theta^-}$, total steps $K$.
\State Initialize $\theta^- \leftarrow \theta$.
\For{$k = 0, \dots, K-1$}
    \State Sample a jet image $x \sim \mathcal{D}$ and noise $z \sim \mathcal{N}(0, \mathbf{I})$.
    \State Get the number of time steps $N(k)$ from a schedule.
    \State Sample a time index $n \sim \mathcal{U}\{1, \dots, N(k)-1\}$.
    \State Get timesteps $t_n$ and $t_{n+1}$ from the discretization schedule (e.g., Karras et al., 2022 \cite{karras2022elucidatingdesignspacediffusionbased}).
    \State Create two noisy images: $x_{n+1} = x + t_{n+1}z$ and $x_n = x + t_n z$.
    \State Compute the model outputs: $y_{n+1} \leftarrow f_\theta(x_{n+1}, t_{n+1})$, \ $y_n \leftarrow \text{stop\_grad}(f_{\theta^-}(x_n, t_n))$.
    \State Compute the consistency loss using a distance metric $d(\cdot, \cdot)$ (e.g., L1, L2, or LPIPS):
    \Statex \hspace{\algorithmicindent} $\mathcal{L} \leftarrow d(y_{n+1}, y_n)$.
    \State Update the online model parameters $\theta$ using a gradient descent step on $\mathcal{L}$.
    \State Update the target model parameters $\theta^-$ using an EMA schedule $\mu(k)$:
    \Statex \hspace{\algorithmicindent} $\theta^- \leftarrow \mu(k)\theta^- + (1 - \mu(k))\theta$.
\EndFor
\end{algorithmic}
\end{algorithm}

\begin{algorithm}[h!]
\caption{Multistep Consistency Sampling for Jet Images}
\label{alg:consistency_sampling}
\begin{algorithmic}[1]
\Statex \textbf{Part 2: Sampling from the Consistency Model}
\State \textbf{Require:} Trained consistency model $f_\theta$, sampling timesteps $\{\tau_i\}_{i=1}^N$ where $T=\tau_1 > \dots > \tau_N = \epsilon$.
\State Sample initial noise $\hat{x}_T \sim \mathcal{N}(0, T^2 \mathbf{I})$.
\State \Comment{\textit{Initial denoising step}}
\State $\bar{x} \leftarrow f_\theta(\hat{x}_T, T)$.
\For{$i = 2, \dots, N$}
    \State Sample noise $z \sim \mathcal{N}(0, \mathbf{I})$.
    \State \Comment{\textit{Add noise back to the current estimate}}
    \State $\hat{x}_{\tau_{i-1}} \leftarrow \bar{x} + \sqrt{\tau_{i-1}^2 - \epsilon^2} z$.
    \State \Comment{\textit{Denoise from the new time step}}
    \State $\bar{x} \leftarrow f_\theta(\hat{x}_{\tau_{i-1}}, \tau_{i-1})$.
\EndFor
\State \textbf{return} $\bar{x}$.
\end{algorithmic}
\end{algorithm}

\begin{figure}[htbp]
    \centering
    \begin{subfigure}[b]{\columnwidth}  
        \centering
        \includegraphics[width=\textwidth]{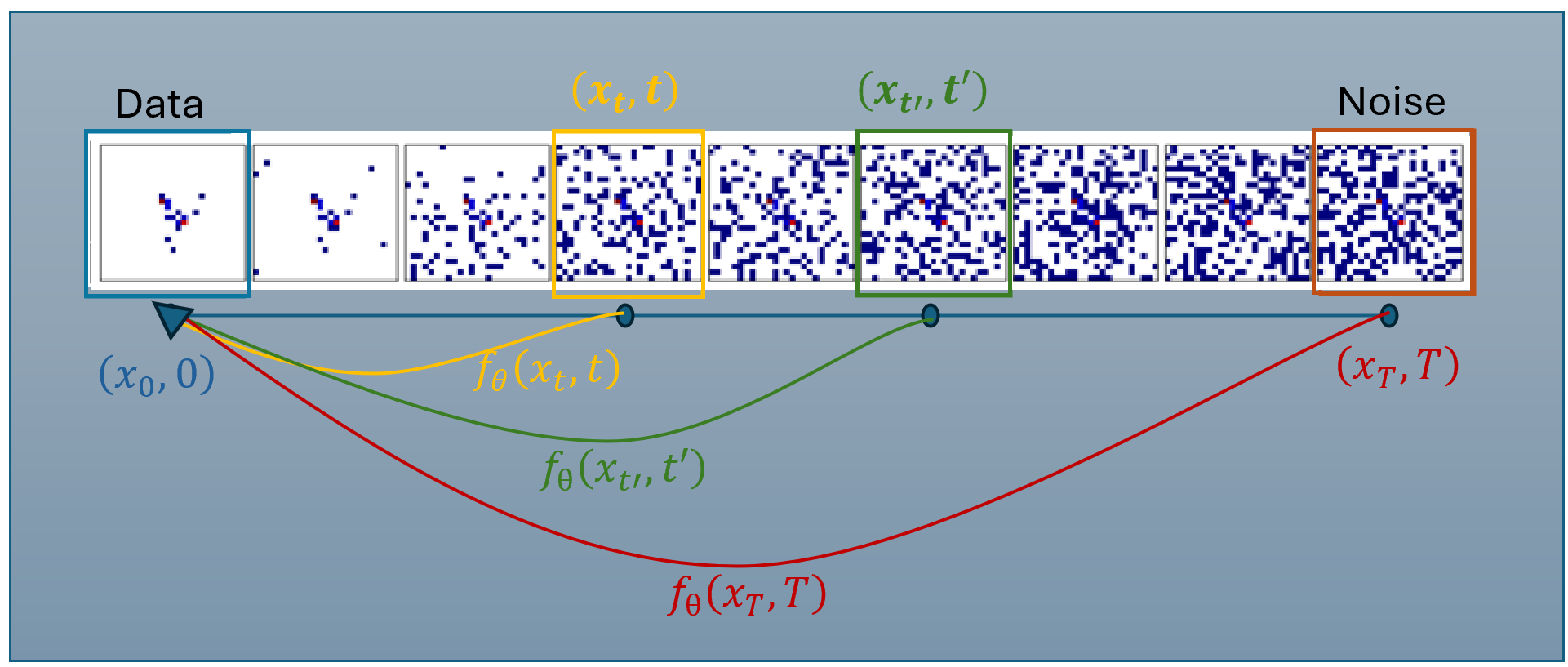}
        \caption{ Mapping consistency from data to noise using multiple timestep trajectories.}
        \label{figconsis2} 
    \end{subfigure}
    \par\medskip  
    \begin{subfigure}[b]{\columnwidth} 
        \centering
        \includegraphics[width=\textwidth]{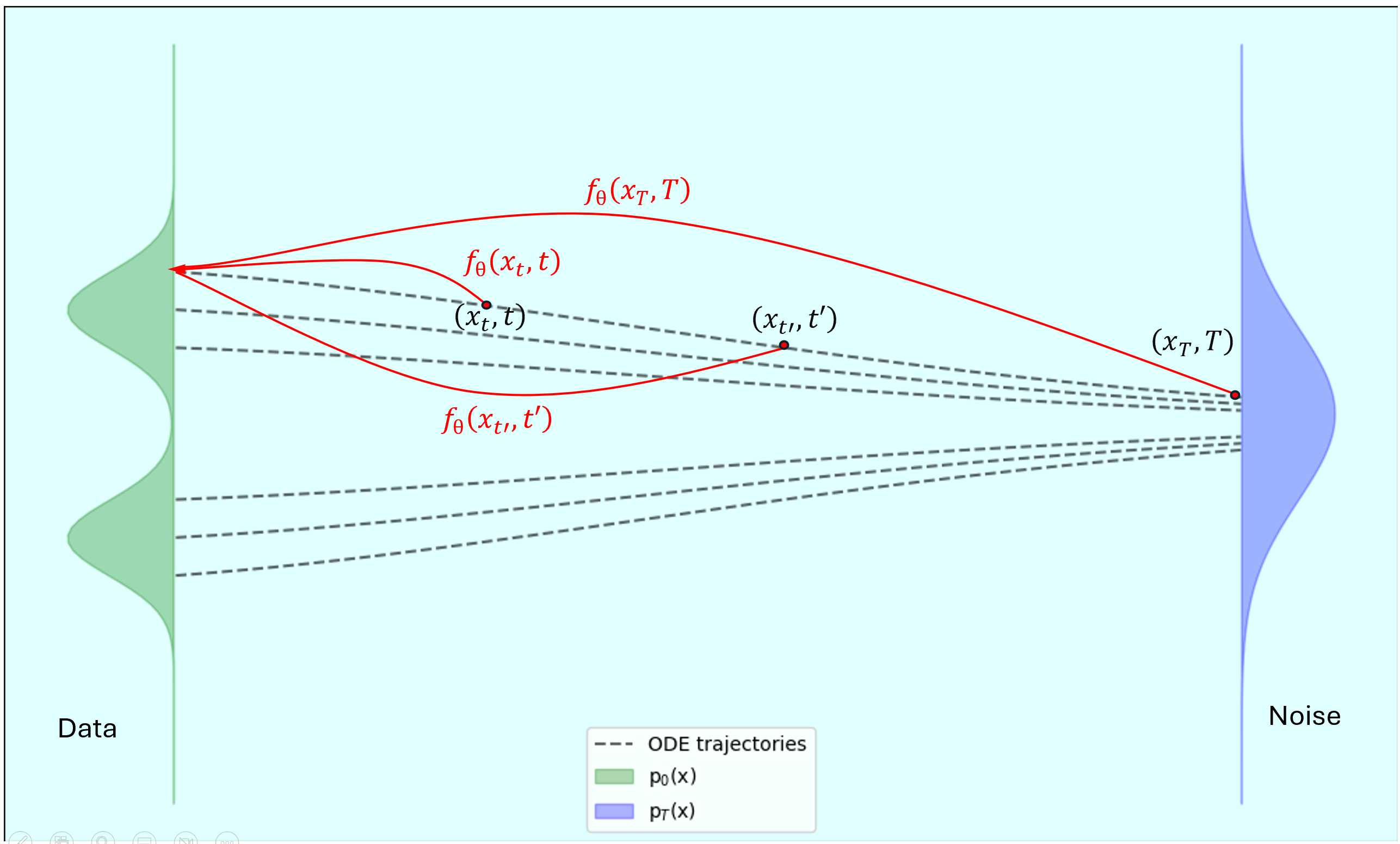}
        \caption{Visual representation of ODE consistent mapping with data and noise distributions.}
        \label{figconsis}
    \end{subfigure}
    \caption{Consistency models are trained to map points along any trajectory of the Probability Flow ODE back to the trajectory’s origin, ensuring that data points maintain consistency across different timesteps and reach the same origin regardless of the chosen path.
}
    \label{figconsis}
\end{figure}

\noindent \textbf{Definition} \quad The probability flow ordinary differential equation (ODE), which emerges as the deterministic counterpart to the stochastic diffusion process, is given by:

\begin{equation}
\frac{d\mathbf{x}_t}{dt} = \boldsymbol{\mu}(\mathbf{x}_t, t) - \frac{1}{2} \sigma(t)^2 \nabla_{\mathbf{x}} \log p_t(\mathbf{x}_t)
\label{eq:pf_ode}
\end{equation}

\noindent where:
\begin{itemize}
  \item \( \mathbf{x}_t \) is the sample at time \( t \),
  \item \( \boldsymbol{\mu}(\mathbf{x}_t, t) \) is the drift term from the forward SDE,
  \item \( \sigma(t) \) is the time-dependent diffusion coefficient,
  \item \( \nabla_{\mathbf{x}} \log p_t(\mathbf{x}_t) \) is the score function.
\end{itemize}

This equation describes the reverse-time dynamics of a diffusion process and underpins the probability flow trajectory followed by samples. It allows sample generation via deterministic ODE solvers rather than stochastic SDE sampling. Given a solution trajectory $\{\mathbf{x}_t\}_{t \in [\epsilon, T]}$ of the PF ODE in Eq. (2), we define the \textit{consistency function} as $f : (\mathbf{x}_t, t) \mapsto \mathbf{x}_{\epsilon}$. A consistency function has the property \textit{of self-consistency}: its outputs are consistent for arbitrary pairs of $(\mathbf{x}_t, t)$ that belong to the same PF ODE trajectory, i.e., $\mathbf{f}(\mathbf{x}_t, t) = \mathbf{f}(\mathbf{x}_{t'}, t')$ for all $t, t' \in [\epsilon, T]$. Formally, this is expressed as:
\begin{equation}
    \mathbf{f}(\mathbf{x}_t, t) = \mathbf{f}(\mathbf{x}_{t'}, t') \quad \forall \ t, t' \in [\epsilon, T].
    \label{eq:consistency}
\end{equation}

As illustrated in Fig. \ref{figconsis}, the goal of a \textit{consistency model}, symbolized as $\mathbf{f}_\theta$, is to estimate this consistency function $\mathbf{f}$ from data by learning to enforce the self-consistency property. This is achieved by enforcing the self-consistency property during training, ensuring that the model's outputs remain stable and accurate across different points in the trajectory \cite{song2023consistency}.
The training of consistency models involves minimizing a loss function that enforces the self-consistency property across various points in the PF ODE trajectory. Let $\mathbf{x}_t$ and $\mathbf{x}_{t'}$ be two points along the same trajectory at times $t$ and $t'$, respectively. The loss function can be defined as:
\begin{equation}
    \mathcal{L}(\theta) = \mathbb{E}_{t, t', \mathbf{x}_t, \mathbf{x}_{t'}} \left[ \| \mathbf{f}_\theta(\mathbf{x}_t, t) - \mathbf{f}_\theta(\mathbf{x}_{t'}, t') \|^2 \right],
\end{equation}
where $\theta$ represents the model parameters, and the expectation is taken over the distribution of trajectories and time points. Algorithm 2 gives the pseudcode for Consistency Training (CT) of original jet image data. The consistency loss function is given in Eq (3) using L2 norm. L1 or Learned Perceptual Image Patch Similarity (LPIPS) norm can also be used. The weight updates are done using Exponential Moving Average (EMA) that keeps model's update parameters steady. Algorithm 3 gives the pseudocode for jet image generation using sampling.

Consistency models have emerged as a powerful class of generative models that support single-step generation at their core while still allowing iterative refinement to balance sample quality and computational cost. Additionally, they enable zero-shot data editing, which is highly valuable for applications requiring flexibility and adaptability without extensive retraining. Consistency models can be trained in two primary modes: \textit{distillation} and \textit{isolation}.

\noindent Training Modes: 
\begin{itemize}
    \item Distillation Mode: In this mode, consistency models distill knowledge from pre-trained diffusion models into a single-step sampler. This approach significantly enhances sample quality compared to traditional distillation methods by leveraging the robust representations learned by diffusion models \cite{song2023consistency}. Moreover, it enables zero-shot image editing applications, where modifications can be made to generated images without additional training.
    \item Isolation Mode: Unlike the distillation mode, isolation mode involves training consistency models independently, without relying on pre-trained diffusion models. This makes consistency models a distinct new class of generative models, capable of learning directly from data to enforce self-consistency properties \cite{song2023consistency}.
\end{itemize}


\section{Jet Image Generation Experiments}
The JetNet dataset consisting of five classes of jets in point cloud representation, the method used to represent them as images which are used for training the two diffusion models: score based and consistency  for generating new jet images,  the jet image generation experimental setup and jet image generation experiments, and the algorithm to reconstruct the jet particle kinematics from genenrated jet image are described below.
\subsection{JetNet Dataset and Jet Image Representation}
JetNet \cite{kansal2021particle} is a synthetic dataset designed to train and evaluate generative models in simulations of particle jets. The JetNet dataset~\cite{kasieczka2021jetnet,jetnetgithub} provides a benchmark for generative models of jet physics. It is a point cloud dataset representing distributions of particles within jets produced in proton-proton collisions with attributes such as transverse momentum and spatial coordinates.  In the JetNet dataset the kinematic information for each particle is stored as $(p_T,\eta, \phi)$ coordinate, where $\phi$ the azimuth angle is computed as the angle of the particle's trajectory with respect to the horizontal $x$ axis and $\eta=-\log(\tan(\theta/2))$ is the pseudorapidity where $\theta$ is the polar angle. In the JetNet dataset, edriniach jet is a collection of 25 to 30 maximum numbers of particles represented as a cloud of points in the kinematic configuration space. In this work, each jet is represented as a single image, where each constituent particle corresponds to a pixel in a 2D rectangular grid. The intensity of the pixels is determined by the relative transverse momentum, $p_T^{rel}=p_T^{\text{particle}}/p_T^{\text{jet}}$, and the location of the pixels is given in the $(\eta^{rel}, \phi^{rel})$ plane defined as $\phi^{\text{rel}}=\phi^{\text{particle}}-\phi^{\text{jet}}\,(\text{mod }2\pi)$ and $\eta^{\text{rel}}=\eta^{\text{particle}}-\eta^{\text{jet}}$. The resulting jet image is a binned histogram in the \( (\eta^{\text{rel}}, \phi^{\text{rel}}) \) space, where the intensity of each bin reflects the \( p_T^{\text{rel}} \) contribution of individual particles, thereby encoding the spatial distribution of energy within the jet. Each jet image is of size 25 x 25 pixels and the y-axis and x-axis are in the range of $\eta^{\text{rel}} \in [-0.4,0.4]$ and $\phi^{\text{rel}}\in [-0.4,0.4]$, respectively. This range is selected so that the jet images are not too sparse. An image representation of jets has the advantage of visualizing the 2D geometry of jets depicting the relative position of particles with respect to each other enabling the visualization of the geometric characteristics such as texture, shape, and jet substructure of each jet in the JetNet dataset.

\subsection{Experimental Setup}
Both score-based models and consistency models were implemented using neural network architectures inspired by publicly available implementations suited for image data. The score-based model follows the configuration from the official Score-SDE implementation \cite{song2020score}, using a U-Net architecture with residual blocks, group normalization, and Swish activation. The consistency model adopts the architecture and training schedule described in \cite{song2023consistency}, with sinusoidal timestep embeddings and multi-scale convolutional layers.

The models were trained using the Adam optimizer with a learning rate of $1 \times 10^{-4}$, batch size of 64, and image resolution of $25 \times 25$ pixels for each of the five jet classes: gluons, quarks, top quarks, W- and Z-bosons, in the JetNet dataset. Training was conducted for 250 epochs. Both diffusion models were implemented in Python 3.9. Table~\ref{tab:hyperparams} summarizes the key hyperparameters for both diffusion models.

\begin{table}[h]
\caption{Hyperparameters used for Score-Based and Consistency Models}
\label{tab:hyperparams}
\centering
\begin{tabular}{|l|c|c|}
\hline
\textbf{Parameter} & \textbf{Score-Based} & \textbf{Consistency Model} \\
\hline
Architecture & U-Net  & U-Net  \\
Activation & Swish & GELU \\
Normalization & GroupNorm (32) & GroupNorm (32) \\
Embedding class & Fourier time & Sinusoidal time \\
Base Channels & 128 & 128 \\
Channel Multiplier & [1, 2, 2, 2] & [1, 2, 2, 2] \\
Residual Blocks per Stage & 2 & 2 \\
Attention Resolutions & None  & None \\
Dropout & 0.1 & 0.1 \\
Image Resolution & $25 \times 25$ & $25 \times 25$ \\
Batch Size & 64 & 64 \\
Optimizer & Adam & Adam \\
Learning Rate & $1 \times 10^{-4}$ & $1 \times 10^{-4}$ \\
EMA Decay & 0.9999 & 0.999 \\
Total Epochs & 250 & 250 \\
Sampling Steps & 15  & 1  or 15  \\
\hline
\end{tabular}
\end{table}

\subsection{Jet Image Generation Experiments}
\begin{figure}[t]
\centering
\includegraphics[width=3.5in]{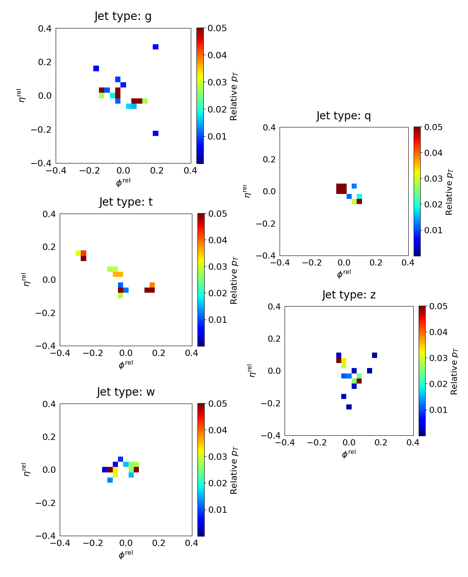}
\caption{Truth jet images of each class: 'gluon', 'top quark, 'W-boson', 'quark' and'Z-boson' from the JetNet dataset. Each image representation corresponds to a single jet event.}
\label{truth}
\end{figure}  

\begin{figure}[t]
\centering
\includegraphics[width=3.5in]{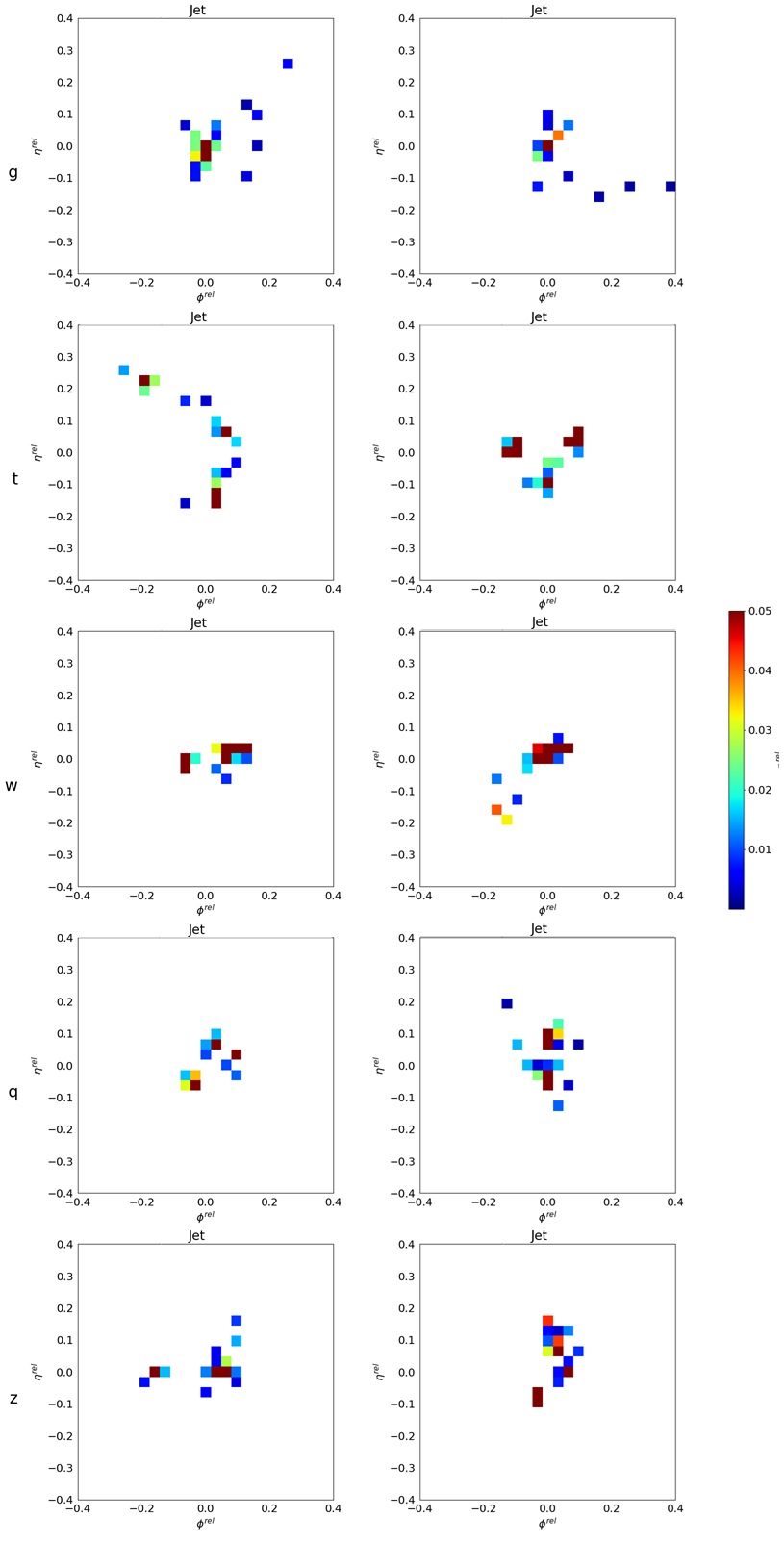}
\caption{Two sample jet images of each class: 'gluon', 'top quark, 'W-boson', 'quark' and 'Z-boson' generated by score-based generative modeling through stochastic differential equations. Each image representation corresponds to a single jet event.}
\label{figsim1}
\end{figure}

Figure \ref{truth} shows one original jet image of each jet class from the JetNet dataset used for training the diffusion models. Figure \ref{figsim1} illustrates the two images for each of the five jet classes, generated using score-based generative modeling through SDEs. The plots y-axis is the relative pseudorapidity (\(\eta^{rel}\)) versus x-axis is the relative azimuthal angle (\(\phi^{rel}\)) for each jet. The left panel shows one generated jet image, while the right panel displays a second sample. The jet images are categorized into five different classes, labeled as 'g', 't', 'w', 'q', and 'z', which stands for gluon, top quark, W-boson, quark, and Z-boson jets, respectively. The color bar on the right denotes the intensity scale, representing the relative transverse momentum of each particle in the \( (\phi^{\text{rel}}, \eta^{\text{rel}}) \) plane. This visualization provides insight into the spatial organization and energy distribution of particles within each jet, highlighting differences and similarities between the jets generated by the diffusion models. The observed clustering patterns and intensity variations reflect underlying kinematic features learned by the model and help characterize jet substructure in the generated data.

\begin{figure}[t]
\centering
\includegraphics[width=3.5in]{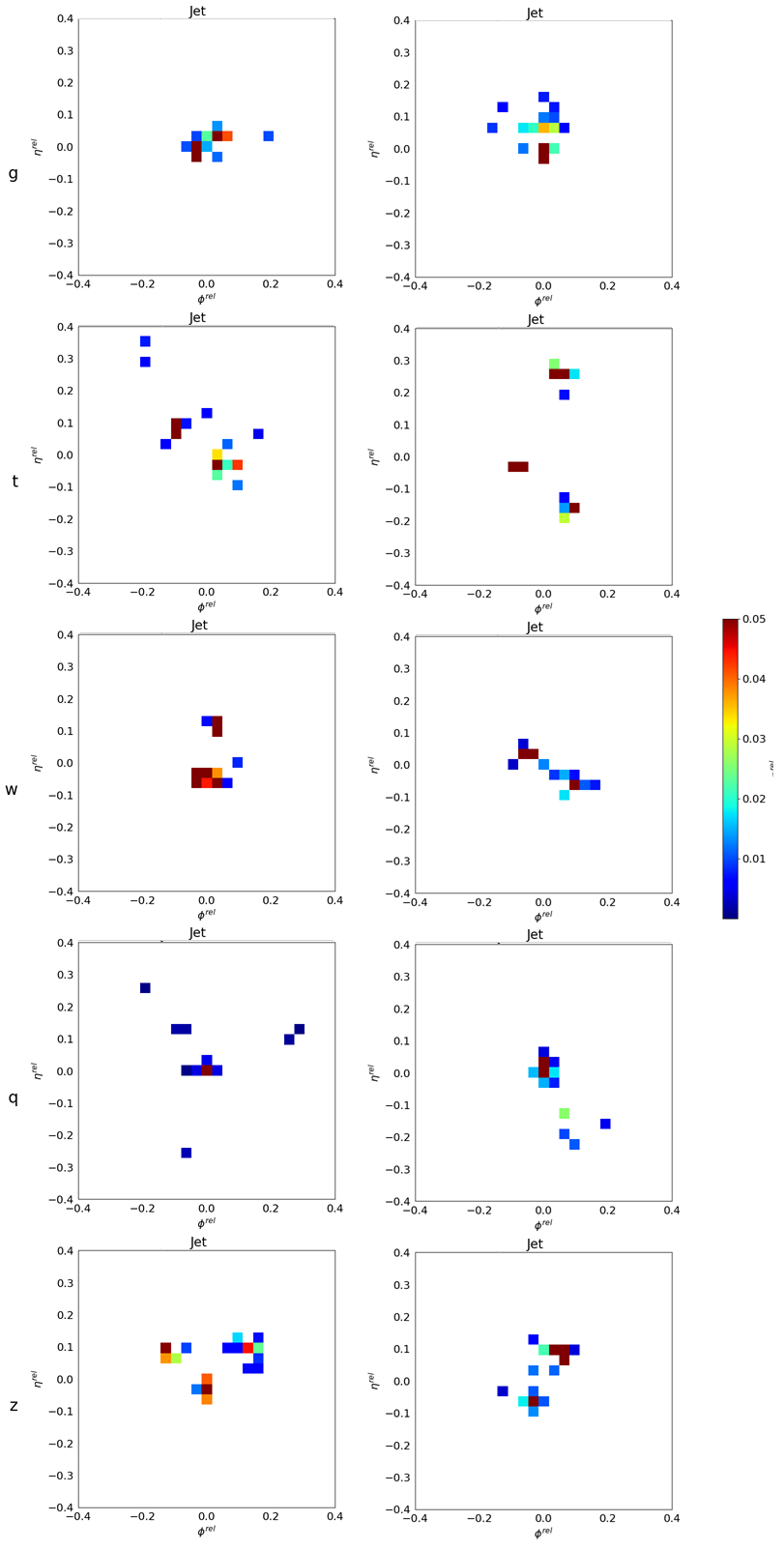}
\caption{Two sample jet images of each class: 'gluon', 'top quark, 'W-boson', 'quark' and 'Z-boson' generated by consistency models. Each image representation corresponds to a single jet event.}
\label{figsim2}
\end{figure}

Figure~\ref{figsim2} presents two sample jet images of each class: 'gluon', 'top quark', 'W-boson', 'quark', and 'Z-boson' generated using the consistency model. These samples illustrate the variability captured by the model across different jet classes and reveal differences in the spatial and angular distribution of jet constituents.

The color gradients represent differences in relative transverse momentum contributions, demonstrating the model's ability to capture fine-grained variations in the spatial distribution of jet constituents. Each image provides a qualitative view of the variation in particle-level structure across different jet classes, illustrating the consistency model's capacity to generate diverse and class-conditional jet images.



\subsection{Reconstructing Particles from Jet Images}

To verify that our diffusion models generate jet images with realistic and diverse jet structures comparable to those in the original JetNet dataset, we develop an algorithm to reconstruct the normalized jet mass from the generated jet constituents. The pseudocode for reconstructing jet kinematic variables \((p_T, \eta, \phi)\) from the generated images is presented in Algorithm 4. Each generated jet image is transformed into a point cloud representation, which is then used to compute the normalized jet mass. This allows for a direct comparison with the normalized jet mass values from the JetNet dataset.

\begin{algorithm}[h]
\caption{Reconstructing Particles from Jet Image(s)}
\label{alg:image_to_particles}
\begin{algorithmic}[1]
  \Require $\texttt{image\_jet}\in \mathbb{R}^{N\times N}\;\lor\;\mathbb{R}^{M\times N\times N}$, $\texttt{maxR}\in\mathbb{R}$
  \Ensure A list of reconstructed jets, each a NumPy array of shape $[\,\text{num\_particles},\,3\,]$ with columns $[\eta,\phi,p_T]$
  \Statex
  \If{ \(\texttt{image\_jet}\) is 2D }
    \State \(\texttt{images} \gets \{\texttt{image\_jet}\}\)  \Comment{Promote single image to a batch of size 1}
  \Else
    \State \(\texttt{images} \gets \texttt{image\_jet}\)
  \EndIf
  \State \(N \gets\) spatial dimension of each square image in \(\texttt{images}\)
  \State \(\Delta \gets \tfrac{2 \,\texttt{maxR}}{N}\)  \Comment{Pixel width in $(\eta,\phi)$ units}
  \State Initialize \(\texttt{jets\_reconstructed} \gets [\,]\)
  
  \ForAll{ \(\texttt{img} \in \texttt{images}\) }
    \State \((\texttt{rows}, \texttt{cols}) \gets \{(i,j)\mid \texttt{img}[i,j] > 0\}\)
    \If{\(\texttt{rows}\) is empty}
      \State Append an empty $(0\times3)$ array to \(\texttt{jets\_reconstructed}\)
      \State \textbf{continue}
    \EndIf
    \State \(\texttt{pt} \gets \texttt{img}[\texttt{rows},\texttt{cols}]\)
    \State \(\eta \gets -\,\texttt{maxR} \;+\; (\texttt{rows} + 0.5)\,\Delta\)
    \State \(\phi \gets -\,\texttt{maxR} \;+\; (\texttt{cols} + 0.5)\,\Delta\)
    \State \(\texttt{particles} \gets \mathrm{stack}([\eta,\,\phi,\,\texttt{pt}],\,\texttt{axis}=-1)\)
    \State Append \(\texttt{particles}\) to \(\texttt{jets\_reconstructed}\)
  \EndFor
  
  \If{original \(\texttt{image\_jet}\) was 2D}
    \State \Return \(\texttt{jets\_reconstructed}[0]\)
  \Else
    \State \Return \(\texttt{jets\_reconstructed}\)
  \EndIf
\end{algorithmic}
\end{algorithm}

\section{Results}
\label{sec:results}

In this section, we present the quantitative analysis of the results obtained by applying score-based models and consistency models for jet image generation. The evaluation metrics used for quantitative analysis are: Fréchet Inception Distance (FID) metric \cite{heusel2017gans}, Wasserstein Distance (WD) \cite{kwon2022scorebasedgenerativemodelingsecretly}, and Diversity Index (DI) \cite{zameshina2023diversediffusionenhancingimage}. 
\subsection{Evaluation metrics}

The FID is calculated as:

\begin{equation}
\text{FID} = \|\mu_r - \mu_g\|^2 + \text{Tr}\left(\Sigma_r + \Sigma_g - 2\left(\Sigma_r \Sigma_g\right)^{\frac{1}{2}}\right)
\end{equation}

Where:
\begin{align*}
\mu_r &= \text{Mean of the original JetNet image features} \\
\Sigma_r &= \text{Covariance matrix of the original JetNet image features} \\
\mu_g &= \text{Mean of the generated image features} \\
\Sigma_g &= \text{Covariance matrix of the generated image features}
\end{align*}

The FID metric \cite{heusel2017gans} is used to quantify the similarity between the particle $p_{T}^{\text{rel}}$ distributions, which are the pixel intensities of the generated jet images and the original JetNet jet images. The FID scores are calculated for 10 runs of each of the models and averaged for each of the five jet classes. Table \ref{tab:fid_scorebased} gives the FID scores for score based generation and Table \ref{tab:fid_consistency} gives the FID scores for consistency based generation, respectively.

\begin{table}[h]
    \caption{FID scores by jet class comparing Score-Based Diffusion Model generated jet images with those of JetNet dataset across epochs.}
    \label{tab:fid_scorebased}
    \centering
    \begin{tabular}{|c||c|c|c|c|c|}
        \hline
        \textbf{Epochs} & $g$ & $t$ & $w$ & $q$ & $z$ \\
        \hline
        10   & 150.2 & 143.1 & 147.0 & 139.8 & 146.3 \\
        20   & 28.3  & 26.9  & 27.6  & 25.8  & 26.9  \\
        30   & 14.5  & 13.2  & 13.8  & 12.7  & 13.5  \\
        40   & 12.4  & 11.0  & 11.5  & 10.7  & 11.3  \\
        60   & 11.6  & 10.2  & 10.7  & 9.9   & 10.6  \\
        80   & 11.3  & 9.9   & 10.3  & 9.5   & 10.2  \\
        100  & 11.2  & 9.8   & 10.2  & 9.4   & 10.1  \\
        120  & 10.8  & 9.5   & 10.0  & 9.0   & 9.7   \\
        140  & 10.5  & \textbf{9.1}   & \textbf{9.6} 
        & \textbf{8.7}   & \textbf{9.3}   \\
         160  & \textbf{10.4}  & 9.3   & 9.9   & 9.0   & 9.4   \\
        250  & 10.7  & 9.8   & 10.5  & 9.2   & 9.9   \\
        \hline
    \end{tabular}
\end{table}

\begin{table}[h]
    \caption{FID scores by jet class comparing Consistency Model generated jet images with those of JetNet dataset across epochs.}
    \label{tab:fid_consistency}
    \centering
    \begin{tabular}{|c||c|c|c|c|c|}
        \hline
        \textbf{Epochs} & $g$ & $t$ & $w$ & $q$ & $z$ \\
        \hline
        10   & 144.6 & 137.2 & 141.5 & 134.9 & 140.1 \\
        20   & 26.2  & 24.8  & 25.5  & 24.3  & 25.3  \\
        30   & 13.3  & 12.0  & 12.5  & 11.8  & 12.3  \\
        40   & 11.1  & 9.8   & 10.3  & 9.5   & 10.1  \\
        60   & 10.3  & 8.9   & 9.4   & 8.6   & 9.2   \\
        80   & 9.8   & 8.2   & 8.9   & 8.0   & 8.6   \\
        100  & 9.5   & 7.9   & 8.6   & 7.8   & 8.3   \\
        120  & 9.0   & 7.3   & 8.1   & 7.1   & 7.8   \\
        140  & \textbf{8.3}   & \textbf{6.7}   & 7.5   & \textbf{6.9}   & \textbf{7.4}   \\
        160  & 8.7   & 7.0   & \textbf{7.4}   & 7.2   & 7.6   \\
        250  & 9.2   & 7.5   & 8.6   & 7.3   & 8.0   \\
        \hline
    \end{tabular}
\end{table}


As shown in Table~\ref{tab:fid_scorebased}, the lowest FID scores for jet image generation using the score-based diffusion model are obtained at 140 epochs for top quark, W-boson, quark, and Z-boson jets. Beyond this point, the FID scores increase, indicating a decline in image quality with further training. For gluon jets, the best FID score is achieved at 160 epochs.

Table~\ref{tab:fid_consistency} shows that the consistency model achieves its lowest FID scores in the range of 6.7 to 8.3 at 140 epochs for gluon, top quark, quark, and Z-boson jets, and at 160 epochs for W-boson jets. In contrast, the score-based model yields higher FID scores, ranging from 8.7 to 10.4.

Overall, the consistency model consistently outperforms the score-based diffusion model across jet classes, achieving significantly lower FID scores and indicating better fidelity to the original JetNet images.

\subsection{Wasserstein Distance }

To complement the FID metric, we computed additional metrics to quantitatively compare the generated jet images with the original images. The Wasserstein Distance (WD) is given by,

\begin{equation}
    W(p, q) = \inf_{\gamma \in \Pi(p, q)} \mathbb{E}_{(x, y) \sim \gamma} [||x - y||]
\end{equation}

Here, $W(p, q)$ represents the Wasserstein distance between two probability distributions $p$ and $q$, while $\Pi(p, q)$ denotes the set of all possible couplings with marginals $p$ and $q$. This metric quantifies the minimum cost required to transform one distribution into the other, effectively capturing the dissimilarity between the generated and original images. The WD shown in Table \ref{tab:additional_metrics} is computed by comparing the distributions of pixel intensities between the original and generated jet images for each of the 10 runs and averaged over these runs for each of the diffusion models.

\subsection{Diversity Index}
The Diversity Index (DI) is formulated as follows.

\begin{equation}
    DI = \frac{1}{N} \sum_{i=1}^N \text{Var}(x_i)
\end{equation}

In this equation, $\text{Var}(x_i)$ represents the variance of feature $x_i$ across the generated samples, in this case, it is the pixel intensity representing the particle $p_{T}^{\text{rel}}$, thereby providing an assessment of the diversity inherent in the generated particle relative transverse momentum. The average DI for the JetNet original image dataset is 0.021. The DI for the generated jet images shown in Table \ref{tab:additional_metrics} is computed for 10 runs and averaged for each of the diffusion models. A higher value of $DI$ has been obtained for the generated jet images using both diffusion models which indicates greater variability and diversity in the generated images.

\begin{table}[h]
    \caption{Wasserstein Distance and Diversity Index Comparing Score-Based and Consistency Models}
    \label{tab:additional_metrics}
    \centering
    \begin{tabular}{lcc}
        \hline 
        \textbf{Model} & \textbf{Wasserstein Distance} & \textbf{Diversity Index} \\
        \hline
        Score-Based Model & 0.35 & 0.78 \\
        Consistency Model & \textbf{0.28} & \textbf{0.82} \\
        \hline
    \end{tabular}
\end{table}

As illustrated in Table \ref{tab:additional_metrics}, the consistency model demonstrates superior performance over the score-based model in terms of both the WD and DI. Specifically, the consistency model yields a lower WD, implying a closer approximation to the original particle (pixel) $p_{T}^{\text{rel}}$ distribution in the generated images, while achieving a higher diversity index, which suggests a broader variety within the generated jet image samples. 

Capturing adequate sample diversity is crucial in generative modeling to prevent mode collapse, which usually happens when generative models produce a limited variety of outputs, effectively "collapsing" onto only a few modes of the data distribution it is supposed to learn\cite{arjovsky2017wgan}. Prior studies ~\cite{krause2021calogan} have emphasized that fast simulation methods must reflect the full range of physical variation, not merely the average behavior. In \cite{butter2022ganplifying}, the authors quantify diversity through variance of physical observables and highlight its role in evaluating generative models and underestimating diversity can lead to undercoverage in downstream inference \cite{nachman2021sbi}. Thus, we interpret the higher DI from the consistency model as a sign of improved mode coverage in a manner that remains physically plausible and we also statistically validate by the statistical analysis presented in Section V A.

These findings collectively indicate that the consistency model not only produces images that is more similar to the original jet image $p_{T}^{\text{rel}}$ distribution but also exhibits greater internal diversity, enhancing its applicability in generating different classes of jet images.

\begin{figure}
    \centering
    \includegraphics[width=0.9\linewidth]{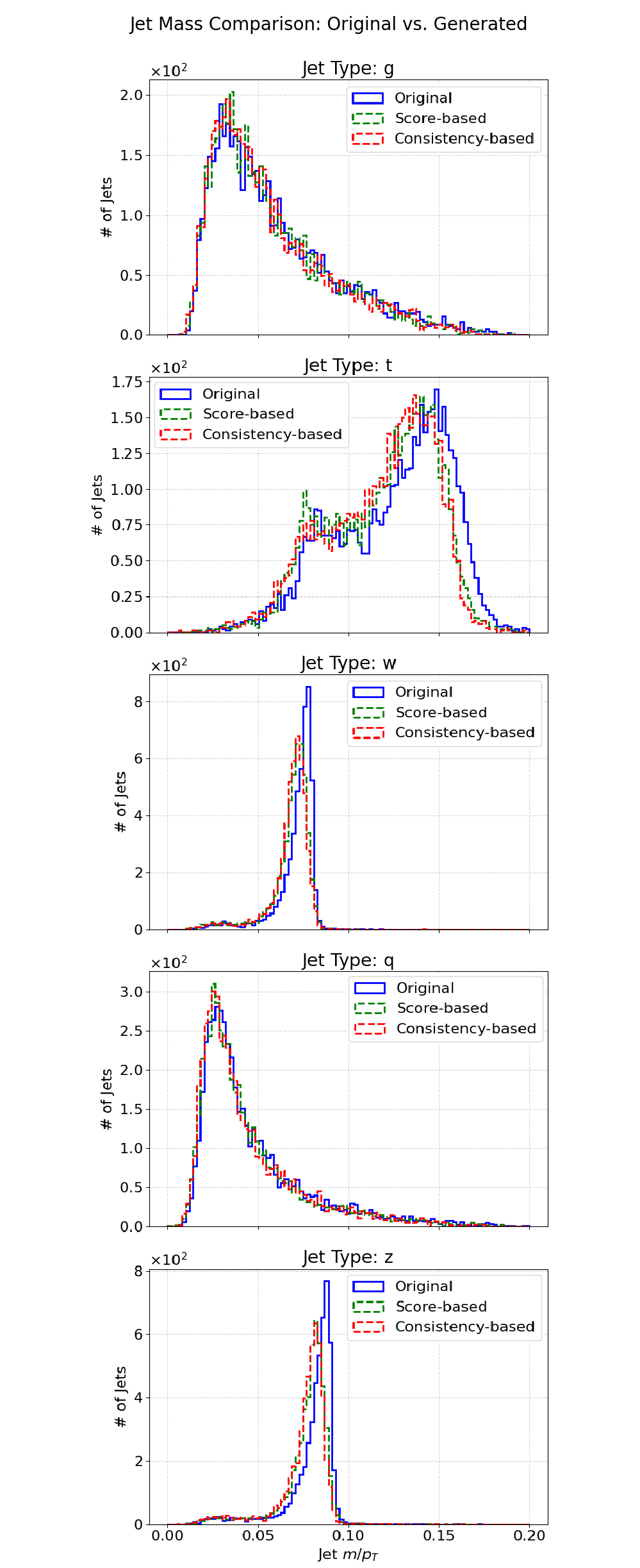}
    \caption{Comparison of normalized jet mass distributions ($m/p_T$) for five jet classes: gluon ($g$), top quark ($t$), $W$ boson ($w$), light quark ($q$), and $Z$ boson ($z$). Each subplot overlays histograms from original JetNet samples (solid blue), score-based diffusion model (dashed green), and consistency-based model (dashed red). The visual alignment of the generated distributions with the original highlights the ability of both generative models to reproduce realistic jet mass profiles, with consistency-based models showing improved agreement in several categories.}
    \label{fig:plots_pt}
\end{figure}

\section{Discussion}

In this section, we analyze the results obtained using the FID, Wasserstein Distance (WD), and Diversity Index (DI) evaluation metrics. We compare the normalized jet mass distributions derived from the generated jet images with those from the original JetNet dataset. Additionally, we present a statistical significance analysis of the results obtained from the two diffusion models used for jet image generation. The strengths and limitations of each model are highlighted, and we discuss the implications of these findings for future research and applications in High Energy Physics (HEP).

The lower FID scores achieved by the consistency models indicate a greater similarity between the distributions of generated and original jet images from the JetNet dataset. This trend is consistent across all five jet classes: gluon, quark, W-boson, Z-boson, and top quark, suggesting that consistency models are particularly effective at capturing low information density structures and fine grained spatial variations that are characteristic of jet images.

Furthermore, the additional evaluation metrics, including Wasserstein Distance and Diversity Index, reinforce these observations. The consistency models not only generate samples that more closely resemble the original jet images, but also show greater variability in generated outputs. This variability may be beneficial in capturing a wider range of jet substructure patterns, which could be relevant for downstream analyses in high energy physics experiments.

Figure~\ref{fig:plots_pt} presents the relative invariant jet mass  normalized by jet transverse momentum  for jets reconstructed from the generated images using both score-based and consistency models. The distributions are compared to those obtained from the original JetNet data. The results show that jets generated using the consistency model more accurately reproduce the normalized jet mass distributions across all five jet classes: gluon, quark, W-boson, Z-boson, and top quark, demonstrating the model’s superior performance in preserving key physical properties.

\subsection{Statistical Significance} \label{sec:significance}

To assess whether the performance differences between the score-based (SB) and consistency (CM) models are statistically significant, we trained and evaluated each model $N=10$ times with different random seeds. For every run we computed Fréchet Inception Distance (FID), Wasserstein Distance ($WD$) and Diversity Index ($DI$). Let
\[
\mu_{\mathrm{SB}},\; \mu_{\mathrm{CM}} \quad\text{and}\quad
\sigma_{\mathrm{SB}}^{2},\; \sigma_{\mathrm{CM}}^{2}
\]
denote the sample means and variances over these runs.

\paragraph*{Two-sample $t$-test.}
Assuming independence between runs, we test the null hypothesis
$H_{0}\!:\, \mu_{\mathrm{SB}} = \mu_{\mathrm{CM}}$ against the
two-sided alternative $H_{1}\!:\, \mu_{\mathrm{SB}} \neq
\mu_{\mathrm{CM}}$. With unequal variances we use Welch’s statistic
\[
t = \frac{\mu_{\mathrm{SB}} - \mu_{\mathrm{CM}}}
         {\sqrt{\frac{\sigma_{\mathrm{SB}}^{2}}{N} + \frac{\sigma_{\mathrm{CM}}^{2}}{N}}},
\]
whose $p$-value is evaluated against a $t$ distribution with
\[
\nu = \frac{\left(\sigma_{\mathrm{SB}}^{2}/N + \sigma_{\mathrm{CM}}^{2}/N\right)^{2}}
         {\frac{\sigma_{\mathrm{SB}}^{4}}{N^{2}(N-1)} +
          \frac{\sigma_{\mathrm{CM}}^{4}}{N^{2}(N-1)}}
\]
degrees of freedom. We adopt a significance level $\alpha = 0.05$.

\paragraph*{Bootstrap confidence intervals.}
Because FID and $WD$ are not guaranteed to be normally distributed,
we also report non-parametric $95\%$ confidence intervals obtained
with the percentile bootstrap ($10,000$ resamples).

\paragraph*{Kolmogorov-Smirnov test on physics observables.}
For each jet class we compare the generated mass distribution
$m/p_{T}$ against original JetNet data using the two-sample
Kolmogorov-Smirnov statistic $D_{\mathrm{KS}}$. The null hypothesis
$H_{0}$ states that both samples arise from the same underlying
distribution. We quote $p$-values and declare significance when
$p<0.01$ (Bonferroni-corrected for the five jet classes).

\begin{table}[h]
\caption{Statistical comparison over $N=10$ runs (mean $\pm$ std).
Bold numbers indicate the better score; ${}^{\dagger}$ denotes
$p<0.05$ under Welch’s test.}
\label{tab:significance}
\centering
\begin{tabular}{@{}lcccc@{}}
\toprule
Metric & \multicolumn{2}{c}{Score-Based} & \multicolumn{2}{c}{Consistency} \\[-1pt]
       & Mean & 95\% CI & Mean & 95\% CI \\ \midrule
FID $\downarrow$ & $9.8 \pm 0.4$  & [9.0, 10.5] & $\mathbf{7.3 \pm 0.3}^{\dagger}$ & [6.7, 7.9] \\
$WD$ $\downarrow$ & $0.35 \pm 0.01$ & [0.33, 0.37] & $\mathbf{0.28 \pm 0.01}^{\dagger}$ & [0.26, 0.30] \\
$DI$ $\uparrow$   & $0.78 \pm 0.02$ & [0.75, 0.81] & $\mathbf{0.82 \pm 0.02}^{\dagger}$ & [0.79, 0.85] \\ \bottomrule
\end{tabular}
\end{table}

All three metrics reject $H_{0}$ in favour of the consistency model
($p<0.01$). Likewise, the KS test shows no statistically significant
difference between CM-generated and real mass distributions for any
jet class ($p>0.05$), whereas the SB model fails this test for the
$g$ and $w$ jets. These results confirm that the observed
performance gap is not due to stochastic fluctuations but reflects a
real improvement introduced by the consistency architecture.

\subsection{\textit{Compute time}}
The score-based diffusion model and the consistency model were executed on an NVIDIA RTX 4070 GPU. Table~\ref{tab:runtime_comparison} reports the runtime of each model for 15 sampling forward passes. As the number of forward passes increases, the consistency model requires less time due to its single-step generation capability. This makes the consistency model highly suitable for fast, large-scale or real-time jet image generation tasks.

\begin{table}[h]
\centering
\caption{Compute Time Comparison Between Models}
\label{tab:runtime_comparison}
\begin{tabular}{>{\bfseries}l|c|c}
\hline
Model & Hardware & Time (ms) \\
\hline
Score-Based & GPU & 18.31 \\
Consistency Model & GPU & \textbf{17.88} \\
\hline
\end{tabular}
\end{table}

\subsection{Advantages of Consistency Models}

The results demonstrate several key advantages of consistency models over score-based models:

\begin{itemize}
    \item Computational Efficiency: The single-step generation capability significantly reduces the time and computational burden associated with iterative refinement processes inherent in diffusion models \cite{song2023consistency}.
    \item Generated Jet Image Quality: As shown by the lower FID, consistency models are able to generate jet images with higher fidelity improving the accuracy of jet image generations. The distillation process allows consistency models to retain high sample quality, often surpassing that of other distilled generative models \cite{song2023consistency}.
    \item Diversity: As shown by the DI comparison of the two diffusion models, the consistency model generated jet images have a higher DI.
    \item Scalability: Training in isolation mode allows consistency models to be scalable and adaptable to different datasets and generative tasks without dependency on pre-trained models \cite{song2023consistency}.
\end{itemize}

\subsection{Limitations and Future Considerations}

Although consistency models show superior performance, there are some limitations that need to be addressed in future research:

\begin{itemize}
    \item Dependence on training data: The quality of the generated images is highly dependent on the quality of the training data. It is crucial to have extensive and representative datasets to train robust models.
    \item Model complexity: Consistency models can be more complex to train due to the need to maintain self-consistency across the Probability Flow ODE trajectories.
    \item Generalization to other datasets: Further evaluation is required to determine how consistency models perform on other jet image generation tasks.
\end{itemize}

Future research could focus on optimizing training algorithms to reduce computational complexity, as well as developing domain-specific metrics that evaluate the physical consistency of the generated jet images. Additionally, integrating prior physical knowledge into the models could further improve the accuracy and interpretability of the generated jet images.

\section{Conclusion}
\label{sec:conclusions}
In this work, we explored jet image generation using two classes of diffusion-based generative models: score-based models and consistency models. Using the JetNet dataset, we evaluated the performance of both models in terms of similarity to the original jet images and normalized jet mass distributions.

Our results show that consistency models consistently outperform score-based models across quantitative metrics such as Fréchet Inception Distance (FID), Wasserstein Distance, and Diversity Index, indicating higher fidelity in the generated images. Comparisons of the normalized jet mass distributions demonstrate that jet images generated by the consistency model closely resemble the original distributions for each of the five jet classes: gluon, quark, W-boson, Z-boson, and top quark. The single-step generation capability of consistency models provides a significant advantage in computational efficiency, making them well-suited for large-scale simulations required in HEP experiments. Statistical significance analyses using the two-sample t-test and Kolmogorov–Smirnov test further confirm the improvements offered by consistency models. These findings suggest that consistency models represent a meaningful advancement over traditional score-based approaches for jet image generation in the context of high-energy physics.


\section*{Acknowledgment}
The authors thank the Artificial Intelligence Imaging Group (aiig) at the University of Puerto Rico, Mayaguez for the computational facility. The authors thank Dr. Jesse Thaler, Professor at the MIT Department of Physics and Director of the Institute of Artificial Intelligence and Fundamental Interactions (IAIFI) and Dr. Shuchin Aeron, Professor, Department of Electrical and Computer Engineering, Tufts for their useful feedback.

\bibliographystyle{IEEEtran}

\def\refname{\vadjust{\vspace*{-1em}}} 

\end{document}